\documentstyle[prb,aps,floats,psfig,oldlfont]{revtex}

\begin{document}
%\twocolumn[
\hsize\textwidth\columnwidth\hsize\csname@twocolumnfalse\endcsname

\title{Continuous quantum measurement of two coupled quantum dots using
a point contact: A quantum trajectory approach}
\author{Hsi-Sheng Goan~${}^1$ \cite{goan}, G.~J.~Milburn~${}^1$,
H.~M.~Wiseman~${}^{2}$, and He Bi Sun~${}^1$}
\address{
${}^1$Center for Quantum Computer Technology and Department of
Physics, The University of Queensland,
Brisbane, Qld 4072 Australia\\
${}^2$School of Science, Griffith University, Nathan, Brisbane,
Qld 4111 Australia}
%\date{e-print: cond-mat/0006333}
\maketitle
\draft

\begin{abstract}
We obtain the finite-temperature unconditional master equation of
the density matrix for two coupled quantum dots (CQD) when one dot is
subjected to a measurement of its electron occupation number using a
point contact (PC).
To determine how the CQD system state depends on the actual current
through the PC device, we use the so-called quantum trajectory method
to derive the zero-temperature conditional master equation.
We first treat the electron tunneling through the PC barrier as a
classical stochastic point process (a quantum-jump model).
Then we show explicitly that our results can be
extended to the quantum-diffusive limit
when the average electron tunneling rate is very large compared to the
extra change of the tunneling rate due to the presence of the electron
in the dot closer to the PC.
We find that in both quantum-jump and quantum-diffusive
cases, the conditional dynamics of the CQD system can be described by
the stochastic Schr\"{o}dinger equations for its conditioned state
vector if and only if the information carried away from the CQD system by
the PC reservoirs can be recovered by the perfect detection of the
measurements.

\end{abstract}

\pacs{85.30.Vw,03.65.Bz,03.67.Lx}

%]

%%%%%%%%%%%%%%%%%%%%%%%%%%%%%%%%%%%%%%%%%%%%%%%%%%%%%%%%%%
\section{Introduction}
%%%%%%%%%%%%%%%%%%%%%%%%%%%%%%%%%%%%%%%%%%%%%%%%%%%%%%%%%%
The origins and mechanisms of decoherence (dephasing) for quantum
systems in condensed matter physics have attracted much attention recently
due to a number of studies in nanostructure mesoscopic
systems \cite{Datta95,Schon97,Buks98,Nakamura99,Sprinzak99}
and various proposals for quantum
computers \cite{Kane98,Loss98,Schoen97,Bonadeo98}.
One of the issues is the connection
between decoherence and quantum
measurement \cite{Wheeler83,Braginsky92}
for a quantum system.
It was reported in a recent experiment\cite{Buks98}  with a
``which-path'' interferometer that Aharonov-Bohm interference is
suppressed owing to the measurement of which path an electron takes
through the double-path interferometer. A biased quantum point contact
(QPC) located close to a quantum dot, which is built in one of the
interferometer's arms, acts as a measurement device. The change of
transmission coefficient of the QPC, which depends on the electron
charge state of the quantum dot, can be detected. The decoherence rate
due to the measurement by the QPC in this experiment has been
calculated in Refs.\
\cite{Aleiner97,Levinson97,Stodolsky98,Buttiker00,Gurvitz97}.

A quantum-mechanical two-state system, coupled to a dissipative
environment, provides a universal model for many physical systems.
The indication of quantum coherence can be regarded as the oscillation
or the interference between the probability amplitudes of finding a
particle between the two states. In this paper, we consider the
problem of an electron tunneling between two coupled quantum dots
(CQDs) using a low-transparency point contact (PC) or tunnel junction
as a detector (environment) measuring the position of the electron
(see Fig.\ \ref{fig:PC}).
This problem has been extensively studied in Refs.\
\cite{Gurvitz97,Gurvitz98,Korotkov99,Korotkov99b,Korotkov00,Makhlin98,Makhlin00,Korotkov00b,Korotkov00c}.
The case of measurements by a general QPC detector with arbitrary transparency
has also been investigated in Refs.\
\cite{Aleiner97,Levinson97,Stodolsky98,Buttiker00,Hackenbroich98,Averin00}.
In addition, a similar system measured by a single electron transistor
rather than a PC has been studied
in Refs.\ \cite{Shnirman98,Makhlin98,Korotkov99b,Makhlin00,Korotkov00c,Wiseman00,Averin00b,Brink00}.
The influence of the detector (environment) on the measured system can
be determined by the reduced density matrix obtained by tracing out
the environmental degrees of the freedom in the total, system plus
environment, density matrix. The master equation (or rate equations)
for this CQD system have been
derived and analyzed
in Refs.\ \cite{Gurvitz97,Stodolsky98}
(here we refer to the rate equations as the first order differential
equations in time for both diagonal and off-diagonal reduced density matrix
elements).
This (unconditional) master equation
is obtained when the results of all
measurement records
(electron current records in this case)
are completely ignored or averaged over, and
describes only the ensemble average property
for the CQD system.
However, if a measurement
is made on the system and the results are available, the state or density
matrix is a
conditional state conditioned on the measurement results.
Hence the deterministic, unconditional master equation cannot
describe the conditional dynamics of the CQD system in a single realization
of continuous measurements which reflects the stochastic nature of an
electron tunneling through the PC barrier.
Consequently, the conditional master equation should be employed.
In condensed matter physics usually many identical quantum systems are
prepared at the same time and a measurement is made upon the systems.
For example, in nuclear or electron magnetic resonance experiments,
generally an ensemble of systems of nuclei and
electrons are probed to obtain the resonance signals.
This implies that the measurement result in this case is
an average response of the ensemble systems.
On the other hand, for various proposed condensed-matter quantum computer
architectures \cite{Kane98,Loss98,Schoen97,Bonadeo98},
how to readout physical properties of a single electronic qubit,
such as charge or spin at a single electron level, is demanding.
This is a non-trivial problem since it involves
an individual quantum particle measured by a practical detector
in a realistic environment. It is particularly important to take account
of the decoherence introduced by the measurements on the qubit
as well as to understand how the quantum state of the qubit,
conditioned on a particular single realization of measurement,
evolves in time for the purpose of quantum computing.

Korotkov \cite{Korotkov99,Korotkov00} has obtained the Langevin rate
equations for the CQD system.
These rate equations describe the random
evolution of the density matrix that both conditions, and is conditioned
by, the
PC detector output.
In his approach, the individual electrons tunneling
through the PC barrier were ignored and the tunneling current was treated
as a continuous,
diffusive variable. More precisely, he considered the change of the
output current average over some small time $\tau$, $\langle
I\rangle$, with respect to the average current $I_i$, as a Gaussian
white noise distribution. He then updated $\langle I\rangle$ in the
density-matrix elements using the new values of $\langle I\rangle$
after each time interval $\tau$. However, treating the tunneling
current as a continuous, diffusive variable is valid only when the
average electron tunneling rate is very large compared to the extra change
of the tunneling rate due to the presence of the electron in the
dot closer to the PC. The resulting derivation of the stochastic rate
equations is
semi-phenomenological,
based on basic physical reasoning to deduce the
properties of the
density matrix elements, rather than microscopic.

To make contact with the
measurement output, in this paper
we present a {\em quantum trajectory}
\cite{Carmichael93,Dalibard92,Gisin92,Wiseman93,Gagen93,Hegerfeldt93,Presilla96,Mensky98,Plenio98,Wiseman00}
measurement analysis to the CQD system.
We first use the quantum open
system approach \cite{Carmichael93,Gardiner91,WallsMilb94,Sun99} to
obtain the unconditional Markovian master equation for the CQD system, taking
into account the finite-temperature effect of the PC reservoirs.
Particularly, we assume the transparency of the PC detector is small,
in the tunnel-junction limit.
Subsequently, we
derive microscopically the zero-temperature conditional master equation
by treating the electron tunneling through the PC
as a classical stochastic point process  (also called a
{\em quantum-jump} model) \cite{Wiseman93,Plenio98,Wiseman00}.
Generally the evolution of the system state undergoing
quantum jumps (or other stochastic processes) is known as a
quantum trajectory \cite{Carmichael93}.
Real measurements (for example the photon number detection) that
correspond approximately to the ideal quantum-jump (or
point-process) measurement are made regularly in experimental
quantum optics. For almost all-infinitesimal time intervals, the
measurement result is null (no photon detected). The system in this
case changes infinitesimally, but not unitarily. The nonunitary component
reflects the
changing probabilities for future events conditioned on past null events.
At randomly
determined times (conditionally Poisson distributed), there is a
detection result. When this occurs, the system undergoes a finite
evolution, called a {\em quantum jump}. In reality these point
processes are not seen exactly due to a finite frequency response of
the circuit that averages each event over some time. Nevertheless, we
first take the zero-response time limit and consider the electron
tunneling current consisting of a sequence of random $\delta$ function
pulses, i.e., a series of stochastic point processes.
Then we show explicitly that our results can be extended to the
quantum-diffusive limit and reproduce the rate equations obtained by Korotkov
\cite{Korotkov99,Korotkov00}. We refer to the case studied by
Korotkov \cite{Korotkov99,Korotkov00} as
quantum diffusion, in contrast to the case of quantum jumps considered here.
Hence our quantum trajectory approach
may be considered as a formal derivation \cite{Korotkov-derivation}
of the rate equations in
Refs.\ \cite{Korotkov99,Korotkov00}.
We find in both quantum-jump and quantum-diffusive cases
that the conditional dynamics of
the CQD system can be described by the
stochastic Schr\"{o}dinger equations
(SSEs) \cite{Carmichael93,Dalibard92,Wiseman93,Presilla96,Plenio98}
for the conditioned state vector, provided that the information
carried away from the CQD system by the PC reservoirs can be recovered by
the perfect detection of the measurements.

This paper is organized as follows.
In Sec.~\ref{sec:masterEq}, we sketch the derivation of
the finite-temperature unconditional master equation for the QCD system.
To determine how the CQD system state depends on the actual current
through the PC device, we derive in Sec.~\ref{sec:jump}
the zero-temperature conditional master equation and the SSE in
the quantum-jump model. Then in Sec.~\ref{sec:diffusive}
we extend the results to the case of quantum diffusion
and obtain the corresponding conditional master equation and SSE.
The analytical results in terms of Bloch sphere variables for
the conditional dynamics are
presented in Sec.~\ref{sec:analytic}.
Specifically, we analyze in this section the localization rate
and mixing rate \cite{Shnirman98,Makhlin98,Makhlin00}.
Finally, a short
conclusion is given in Sec.~\ref{sec:conclusion}.
Appendix \ref{sec:equivalence} is devoted to the demonstration of the
equivalence between the conditional stochastic rate
equations in Refs.\ \cite{Korotkov99,Korotkov99b,Korotkov00}
and those derived microscopically in the present paper.

%%%%%%%%%%%%%%%%%%%%%%%%%%%%%%%%%%%%%%%%%%%%%%%%%%%%%%%%%%
\section{unconditional master equation for the CQD and PC model}
\label{sec:masterEq}
%%%%%%%%%%%%%%%%%%%%%%%%%%%%%%%%%%%%%%%%%%%%%%%%%%%%%%%%%%

The appropriate way to approach quantum measurement problems is
to treat the measured system, the detector
(environment), and the coupling between them microscopically.
Following from Refs.\ \cite{Gurvitz97,Korotkov99,Korotkov00},
we describe the whole system (see Fig.\ \ref{fig:PC})
by the following Hamiltonian:
\begin{equation}
{\cal H}= {\cal H}_{CQD}+{\cal H}_{PC}+{\cal H}_{coup}
\end{equation}
where
\begin{eqnarray}
{\cal H}_{CQD}&=&\hbar\left[ \omega_1 c_1^\dagger c_1
+\omega_2c_2^\dagger c_2
+\Omega(c_1^\dagger c_2+ c_2^\dagger c_1)\right],
\label{HCQD} \\
{\cal H}_{PC}&=&\hbar \sum_k
\left(\omega_k^L a_{Lk}^\dagger a_{Lk}
+\omega_k^R a_{Rk}^\dagger a_{Rk}\right)
+ \sum_{k,q}
\left(T_{kq}a_{Lk}^\dagger a_{Rq}+ T^*_{qk}a_{Rq}^\dagger a_{Lk} \right),
\label{HPC} \\
{\cal H}_{coup}&=&\sum_{k,q} c_1^\dagger c_1
\left(\chi_{kq}a_{Lk}^\dagger a_{Rq}
+ \chi^*_{qk}a_{Rq}^\dagger a_{Lk} \right).
\label{Hcoup}
\end{eqnarray}
%Here $\hbar=h/(2\pi)$ with the Planck constant $h$.
${\cal H}_{CQD}$ represents the effective tunneling Hamiltonian
for the measured CQD system.
For simplicity, we assume strong inner and inter dot Coulomb
repulsion, so only one electron can occupy this CQD system. We label
each dot with an index $1,2$ (see Fig.\ \ref{fig:PC})
and let $c_i$  ($c_i^\dagger$) and
$\hbar\omega_i$ represent the electron annihilation (creation)
operator and energy for a single electron state in each dot
respectively. The coupling between these two dots is given by
$\hbar\Omega$. The tunneling Hamiltonian for the PC
detector is represented by ${\cal H}_{PC}$ where $a_{Lk}$, $a_{Rk}$
and $\hbar\omega_k^L$, $\hbar\omega_k^R$ are respectively the fermion
(electron) field annihilation operators and energies for the left and
right reservoir states at wave number $k$. One should not be confused by
the electron in the CQD with the electrons in the PC reservoirs. The
tunneling matrix element between states $k$ and $q$ in left and right
reservoir respectively is given by $T_{kq}$. Eq. (\ref{Hcoup}), ${\cal
H}_{coup}$, describes the interaction between the detector and the
measured system, depending on which dot is occupied. When the electron
in the CQD system is close by to the PC (i.e., dot $1$ is occupied),
there is a change in the PC tunneling barrier. This barrier change
results in a
change of the effective tunneling amplitude from $T_{kq}\rightarrow
T_{kq}+\chi_{kq}$. As a consequence, the current through the PC is
also modified. This changed current can be detected,
and thus a measurement of the location of the electron in the CQD system is
effected.

The total density operator $R(t)$ for the entire system in the
interaction picture satisfies:
\begin{equation}
\dot{R}_I(t)=-\frac{i}{\hbar}\left[H_I(t),R_I(0)\right]
-\frac{1}{\hbar^2}\int_0^t dt' \left[H_I(t), [H_I(t'),R_I(t')]\right].
\label{RIint-diff}
\end{equation}
%where $\dot{R}=dR/dt$.
The dynamics of the entire system is determined by the
time-dependent Hamiltonian \cite{smallOmega}:
\begin{eqnarray}
H_{I}(t)=\sum_{k,q} \left(T_{kq}
+\chi_{kq} c_1^\dagger c_1 \right) a_{Lk}^\dagger a_{Rq}
e^{i(\omega_k^L-\omega_k^R)t}+H.C.,
\label{HI}
\end{eqnarray}
where we have treated the sum of the tunneling Hamiltonian parts
in ${\cal H}_{PC}$
and ${\cal H}_{coup}$ as the interaction Hamiltonian ${\cal H}_{I}$, and
$H.C.$ stands for Hermitian conjugate of the entire previous
term.
By tracing both sides of Eq.\
(\ref{RIint-diff}) over the bath (reservoir) variables
and then changing from the
interacting picture to the Schr\"{o}dinger picture,
we obtain \cite{Carmichael93,Gardiner91,WallsMilb94} the
finite-temperature, Markovian master equation for the CQD system:
\begin{equation}
\dot{\rho}(t)=-\frac{i}{\hbar}[{\cal H}_{CQD}, \rho(t)]
+{\cal D}[{\cal T_+}+{\cal X_+} n_1]\rho(t)
+{\cal D}[{\cal T}_-^*+{\cal X}_-^* n_1]\rho(t),
\label{masterEqT}
\end{equation}
where $\rho(t)={\rm Tr}_B{R(t)}$ and Tr$_B$ indicates
a trace over reservoir variables.
In arriving at Eq.\ (\ref{masterEqT}), we have made the following
assumptions and approximations:
(a)treating the left and right
fermion reservoirs in the PC as
thermal equilibrium free electron baths,
(b)weak system-bath coupling, (c)small transparency of the PC, i.e.,
in the tunnel-junction limit, (d)uncorrelated and factorizable
system-bath initial condition
(e)relaxation time scales of the reservoirs being much shorter
than that of the system state,
(f)Markovian approximation,
(g)$|eV|, k_B T \ll \mu_{L(R)}$,
and (h)energy-independent electron tunneling amplitudes
and density of states over the bandwidth of $\max(|eV|,k_BT)$.
Here $k_B$ is the Boltzmann constant, $T$ represents the temperature,
$eV=\mu_L-\mu_R$ is the external bias applied across the PC,
and $\mu_L$ and $\mu_R$ stand for the chemical potentials in the
left and right reservoirs respectively.
In Eq.\ (\ref{masterEqT}), $n_1=c_1^\dagger c_1$ is the 
occupation number operator for dot 1.
The parameters ${\cal T_\pm}$ and ${\cal X_\pm}$
are given by
\begin{mathletters}
\label{rate}
\begin{eqnarray}
|{\cal T}_\pm|^2&=&D_\pm= 2\pi e |T_{00}|^2 g_L g_R V_\pm/\hbar,
\label{D}
\\
|{\cal T}_\pm+{\cal X}_\pm|^2&=&D'_\pm
= 2\pi e |T_{00}+\chi_{00}|^2 g_L g_R V_\pm/\hbar,
\label{D'}
\end{eqnarray}
\end{mathletters}
where $D_\pm$ and $D'_\pm$ are the average electron tunneling rates
through the PC barrier in positive and negative bias directions
at finite temperatures,
without and with the presence of the electron in dot 1 respectively.
Here the effective finite-temperature external bias
potential, $ eV_{\pm}$ is given by the following expression:
\begin{equation}
eV_\pm \equiv \frac{\pm eV} {1-\exp[\mp eV/(k_BT)]}.
\label{Vpm}
\end{equation}
$T_{00}$ and $\chi_{00}$ are energy-independent tunneling amplitudes
near the average chemical potential, and $g_L$ and $g_R$ are the
energy-independent density of states for the left and right fermion baths.
Note that the average electron currents through the PC barrier is
proportional to the difference between the average electron
tunneling rate in opposite directions.
Hence, the average currents $eD=e(D_+ -D_-)$ and $eD'=e(D'_+ -D'_-)$,
following from  Eq.\ (\ref{rate}) and (\ref{Vpm}),
are temperature independent \cite{Ingold92,Averin91} at least
for a range of low temperatures $k_B T\ll \mu_{L(R)}$.
In addition, the current-voltage characteristic in the linear response
region $|eV| \ll \mu_{L(R)}$ is of the same form as for an Ohmic resistor,
though the nature of charge transport is quite different in both cases.

We have also introduced, in Eq.\ (\ref{masterEqT}), an elegant superoperator
\cite{Wiseman93,Wiseman00,Wiseman93b,Wiseman94,Wiseman96} ${\cal D}$,
widely used in measurement
theory in quantum optics. Physically the ``irreversible'' part
caused by the influence
of the environment in the unconditional master equation,
is represented by the ${\cal D}$ superoperator.
Generally superoperators transform one operator into another
operator.
Mathematically, the expression ${\cal D}[B]\rho$ means that
superoperator ${\cal D}$ takes its operator argument $B$, acting on
$\rho$. Its precise definition is in terms of another two
superoperators ${\cal J}$ and ${\cal A}$:
\begin{equation}
{\cal D}[B]\rho={\cal J}[B]\rho - {\cal A}[B]\rho,
\label{defcalD}
\end{equation}
where
\begin{eqnarray}
{\cal J}[B]\rho &=& B \rho B^\dagger,
\label{defcalJ}\\
{\cal A}[B]\rho &=& (B^\dagger B \rho +\rho B^\dagger B)/2.
\label{defcalA}
\end{eqnarray}
The form of the master equation (\ref{masterEqT}), defined through
the superoperator ${\cal D}[B]\rho(t)$, preserves the positivity of the
density matrix operator $\rho(t)$. Such a Markovian master equation
is called a Lindblad \cite{Lindblad73} form.

To demonstrate the equivalence between the master equation
(\ref{masterEqT}) and the rate equations derived in Ref.\
\cite{Gurvitz97}, we evaluate the density matrix operator
in the same basis as in Ref.\ \cite{Gurvitz97} and obtain
\begin{mathletters}
\label{rateEq}
\begin{eqnarray}
\dot\rho_{aa}(t)& = &i\Omega[\rho_{ab}(t)-\rho_{ba}(t)]\;,
\label{rateEqa}\\
\dot\rho_{ab}(t)& = & i\varepsilon\rho_{ab}(t)+i\Omega[\rho_{aa}(t)
-\rho_{bb}(t)]-(|{\cal X}_T|^2/2)\rho_{ab}(t)
+i\, {\rm Im} ({\cal T}_+^* {\cal X}_+ -{\cal T}_-^* {\cal X}_-)\rho_{ab}(t)
\label{rateEqc}
\end{eqnarray}
\end{mathletters}
Here $\hbar \varepsilon=\hbar (\omega_2-\omega_1)$ is the energy
mismatch between the two dots,
$\rho_{ij}(t)=\langle i|\rho(t)| j\rangle$,
and $\rho_{aa}(t)$ and $\rho_{bb}(t)$ are the probabilities
of finding the electron in dot 1 and dot 2 respectively.
The rate equations for the other two density matrix elements can be
easily obtained from the relations: $\rho_{bb}(t)=1-\rho_{aa}(t)$ and
$\rho_{ba}(t)=\rho_{ab}^*(t)$.
Compared to an isolated CQD system, the presence of
the PC detector introduces two effects to the CQD system.
First,  the imaginary part of 
$({\cal T}_+^* {\cal X}_+ -{\cal T}_-^* {\cal X}_-)$
(the last term in Eq.\ (\ref{rateEqc})) causes an effective
temperature-independent shift in the
energy mismatch between the two dots.
%Here, ${\cal T}={\cal T}_+ - {\cal T}_-$ and
%${\cal X}={\cal X}_+ - {\cal X}_-$ are temperature-independent quantities.
Here,
$({\cal T}_+^* {\cal X}_+ -{\cal T}_-^* {\cal X}_-)={\cal T}^* {\cal X}$
is a temperature-independent quantity,
where ${\cal T}={\cal T}_+(0)$,  ${\cal X}={\cal X}_+(0)$;
i.e., ${\cal T}_+$ and ${\cal X}_+$ evaluated at zero 
temperature respectively.
Second, it generates a decoherence
(dephasing) rate
\begin{equation}
\Gamma_d={|{\cal X}_T|^2}/{2}
\label{dephaserate}
\end{equation}
for the off-diagonal density matrix elements,
where $|{\cal X}_T|^2=|{\cal X}_+|^2 + |{\cal X}_-|^2$.
We note that the decoherence
rate comes entirely from the effect of the measurement revealing where
the electron in the CQDs is located. If the PC detector does not
distinguish which of the dots the electron occupies, i.e.,
${\cal X}_\pm=0$,
then $\Gamma_d=0$.
The rate equations in Eq.\ (\ref{rateEq})
are exactly the same as the zero-temperature rate
equations in Ref.\ \cite{Gurvitz97} if we assume
that the tunneling amplitudes are real,
$T_{00}=T^*_{00}$ and $\chi_{00}=\chi^*_{00}$.
In that case, the last term in Eq.\ (\ref{rateEqc}) vanishes
and $\Gamma_d={\cal X}^2/{2}=(\sqrt{D'}-\sqrt{D})^2/{2}$.
Actually, the relative phase between the two complex tunneling
amplitudes may produce additional effects on
conditional dynamics of the CQD system as well.
This will be shown later when we discuss conditional dynamics.
Physically, the presence of the electron in dot 1
raises the effective tunneling barrier of the PC
due to electrostatic repulsion.
As a consequence,
the effective tunneling amplitude becomes lower, i.e.,
$D'=|{\cal T}+{\cal X}|^2<D=|{\cal T}|^2$.
This sets a condition on the relative phase $\theta$
between ${\cal X}$ and ${\cal T}$:
$\cos\theta<-|{\cal X}|/(2|{\cal T}|)$.

The dynamics of the unconditional rate
equations at zero temperature was analyzed in
Ref.\ \cite{Gurvitz97}.
Here, following from Eqs.\ (\ref{dephaserate}), (\ref{rate}) and
(\ref{Vpm}), we find that the temperature-dependent decoherence rate
due to the PC thermal reservoirs has the following expression:
\begin{equation}
\frac{\Gamma_d(T)}{\Gamma_d(0)}=\frac{e(V_++V_-)}{eV}
=\coth\left(\frac{eV}{2k_B T}\right).
\label{dephaseTemp}
\end{equation}
As expected, $\Gamma_d(T)$ increases with increasing temperature,
although the average tunneling current through the PC
is temperature independent \cite{Ingold92,Averin91}
for the same range of low temperatures $k_B T\ll \mu_{L(R)}$.
This temperature dependence of the decoherence rate is in fact
just the temperature dependence of the zero-frequency
noise power spectrum of the current fluctuation in
a low-transparency PC or tunnel junction \cite{Larkin68}.
The CQD system weakly coupled to another finite-temperature environment
beside the PC detector was discussed in Ref.\
\cite{Korotkov00}. However, the influence of the finite-temperature
PC reservoirs on the CQD system, presented here, was not taken into account.
The finite-temperature decoherence rate of an one-electron state
in a quantum dot due to charge fluctuation of a general QPC
has been calculated in Ref.\ \cite{Levinson97}.
In Ref.\ \cite{Averin00}, the temperature-dependent
decoherence rate for a two-state system caused by a
QPC detector has been discussed
specifically in the context of the measurement problem.

%%%%%%%%%%%%%%%%%%%%%%%%%%%%%%%%%%%%%%%%%%%%%%%%%%%%%%%%%%%%%%%%
\section{Quantum-jump, conditional master equation}
\label{sec:jump}
%%%%%%%%%%%%%%%%%%%%%%%%%%%%%%%%%%%%%%%%%%%%%%%%%%%%%%%%%%%%%%%%
So far we have considered the evolution of reduced density matrix
when all the measurement results are ignored, or averaged over.
To make contact with a single realization of the measurement records
and study the stochastic evolution of the quantum state, conditioned on a
particular measurement realization, we derive in this section the
quantum-jump, conditional master equation at zero temperature.

The nature of the measurable quantities, such as accumulated number of
electrons tunneling through the PC barrier, is stochastic.
On average of course the same current flows in
both reservoirs. However, the current is actually made up of
contributions from random pulses in each reservoir, which do not
necessarily occur at the same time. They are indeed separated in time
by the times at which the electrons tunnel through the PC.
In this section, we treat the electron tunneling current consisting of
a sequence of random $\delta$ function pulses. In other words, the measured
current is regarded as a series of point processes
(a quantum-jump model) \cite{Wiseman93,Plenio98,Wiseman00}.
The case of quantum diffusion will be analyzed in Sec.~\ref{sec:diffusive}.

Before going directly to the derivation, we discuss some general ideas
concerning quantum measurements. If the system under observation is in
a pure quantum state at the beginning of the measurement, then it will
still be in a pure conditional state after the measurement, conditioned on the
result, provided no information is lost. For example, if the initial
normalized state is
$|\psi(t)\rangle$, the unnormalized final state given the result
$\alpha$ at the end of the time interval $[t,t+dt)$ of the
measurement becomes
\begin{equation}
|\tilde{\psi}_\alpha (t+dt)\rangle=M_\alpha (dt)|\psi(t)\rangle,
\label{tildepsi}
\end{equation}
where $\{M_\alpha (t)\}$ represents a set of operators which define
the measurements and satisfies the completeness condition
\begin{equation}
\sum_\alpha M^\dagger_\alpha (t) M_\alpha (t)=1.
\label{Ma}
\end{equation}
Eq.\ (\ref{Ma}) is simply a statement of conservation of probability.
The corresponding unnormalized density matrix, following
from Eq.\ (\ref{tildepsi}), is given by
\begin{equation}
\tilde{\rho}_\alpha (t+dt)
=|\tilde{\psi}_\alpha (t+dt)\rangle \langle\tilde{\psi}_\alpha (t+dt)|
={\cal J}[M_\alpha (dt)]\rho(t),
\label{tilderho}
\end{equation}
where $\rho(t)=|\psi(t)\rangle \langle \psi(t)|$
and the superoperator ${\cal J}$ is defined in Eq.\ (\ref{defcalJ}).
Of course, if the measurement is made but the result is ignored, the
final state will not be pure but a mixture of the possible outcome
weighted by their probabilities.
Consequently, the unconditional density matrix
can be written as
\begin{equation}
\rho(t+dt)= \sum_\alpha \tilde{\rho}_\alpha (t+dt)
=\sum_\alpha {\rm Pr}[\alpha]\rho_\alpha (t+dt),
\label{measurerho}
\end{equation}
where ${\rm Pr}[\alpha]={\rm Tr}[\tilde{\rho}_\alpha (t+dt)]$ stands
for the probability for the system to be observed in the state
$\alpha$, and
$\rho_\alpha (t+dt)= \tilde{\rho}_\alpha (t+dt)/{\rm Pr}[\alpha]$ is
the normalized density matrix at time $t+dt$.

Now we proceed to derive the quantum-jump, conditional master equation
in the following.
Only two measurement operators $M_\alpha (dt)$ for $\alpha=0,1$ are
needed for a measurement record which is a point process. For most
of the infinitesimal time intervals, the measurement result is
$\alpha=0$, regarded as a {\em null} result. On the other hand, at randomly
determined times, there is a result $\alpha=1$, referred as a
{\em detection} of an electron tunneling through the PC barrier.
Formally, we can write the current through the PC as
\begin{equation}
i(t) = e\, {dN(t)}/{dt},
\label{current}
\end{equation}
where $e$ is the electronic charge and $dN(t)$ is a classical point
process which represents the number (either zero or one) of tunneling
events seen in an infinitesimal time $dt$.
We can think of $dN(t)$ as the increment in the number of electrons $N(t)$
in the drain in time $dt$.
It is this variable, the accumulated
electron number transmitted through the PC,
which is used in Refs.\ \cite{Gurvitz97,Shnirman98,Makhlin00}.
The point process is formally defined by the
conditions on the classical random variable $dN_c(t)$:
\begin{eqnarray}
[dN_c(t)]^{2} &=& dN_c(t),  \label{dN} \\
E[dN_c(t)] &=&{\rm Tr}[\tilde{\rho}_{1c} (t+dt)]
={\rm Tr} \{{\cal J}[M_1(dt)]\rho_c(t)\}={\cal P}_{1c}(t) dt.
\label{dN1}
\end{eqnarray}
Here we explicitly use the subscript $c$ to indicate that the quantity
to which it is attached is conditioned on previous measurement
results, the occurrences (detection records) of
the electrons tunneling
through the PC barrier in the past. $E[Y]$ denotes an ensemble
average of a classical stochastic process $Y$. Eq.\ (\ref{dN}) simply
states that $dN_c(t)$ equals either zero or one, which is why it is
called a point process. Eq.\ (\ref{dN1}) indicates that the ensemble
average of $dN_c(t)$ equals the probability (quantum average) of
detecting electrons tunneling through the PC barrier in time $dt$.
In addition,
$dN_c(t)$ is of order $dt$ and obviously all moments (powers) of
$dN_c(t)$ are of the same order as $dt$. Note here that the density matrix
$\rho_c(t)$ is not the solution of the unconditional reduced master
equation, Eq.\ (\ref{masterEq}). It is actually conditioned
by $dN_c(t')$ for $t'<t$.

The stochastic conditional density matrix at a later time $t+dt$ can
be written as:
\begin{equation}
\rho_{c}(t+dt) = dN_c(t) \frac{\tilde{\rho}_{1c}(t+dt)}
{{\rm Tr}[\tilde{\rho}_{1c}(t+dt)]}
+ [1-dN_c(t)]\frac{\tilde{\rho}_{0c}(t+dt)}
{{\rm Tr}[\tilde{\rho}_{0c}(t+dt)]}.
\label{rhoc}
\end{equation}
Eq.\ (\ref{rhoc}) states that when $dN_c(t)=0$ (a null
result), the system changes infinitesimally via the operator $M_0 (dt)$
and hence the $\rho_c(t+dt)=\rho_{0c}(t+dt)$. Conversely, if $
dN_c(t)=1$ (a detection), the system goes through a finite evolution
induced by the operator $M_1 (dt)$, called a {\em quantum jump}. The
corresponding normalized conditional density matrix then becomes
$\rho_{1c}(t+dt)$.
One can see, with the help of Eqs.\ (\ref{current}), that in this
approach the instantaneous system state conditions the measured
current (see Eq.\ (\ref{dN1})) while the measured current itself
conditions the future evolution of the measured system (see Eq.\
(\ref{rhoc})) in a self-consistent manner. It is straightforward to
show that the ensemble average of the conditional density matrix
equals the unconditional one, $E[\rho_c(t)]=\rho(t)$. Tracing over
both sides of Eq.\ (\ref{measurerho}) for $\alpha=0,1$, we obtain
\begin{equation}
{\rm Tr}[\tilde{\rho}_{0c}(t+dt)]
=1-{\rm Tr}[\tilde{\rho}_{1c}(t+dt)].
\label{Trtilderho0}
\end{equation}
Then taking the ensemble average over the stochastic variables
$dN_c(t)$ on both sides of Eq.\ (\ref{rhoc}), replacing $E[dN_c(t)]$
by using Eq.\ (\ref{dN1}), and comparing the resultant equation with
Eq.\ (\ref{measurerho}) completes the verification.

Next we find the specific expression of $\tilde{\rho}_{1c}(t+dt)$ and
$\tilde{\rho}_{0c}(t+dt)$ and derive the conditional master equation
for the CQD system measured by the PC.
If a perfect PC detector  (or efficient measurement) is assumed,
then whenever an electron
tunnels through the barrier, there is a measurement record
corresponding to the occurrence of that event; there are no `misses' in the
count of the electron number. As a result, the information lost
from the system to the reservoirs can be recovered 
using a perfect detector.
Here we assume a zero-temperature case for the efficient measurement.
At finite temperatures, the electrons can, in principle,
tunnel through the PC barrier in both directions.
But experimentally the detector might not be able to
detect these electron tunneling processes
on both sides of the PC barrier. This may result in
information loss at finite temperatures.
Hence, at zero temperature the unconditional master equation
(\ref{masterEqT}) reduces to
\begin{mathletters}
\label{masterEquation}
\begin{eqnarray}
\dot{\rho}(t)&=&-\frac{i}{\hbar}[{\cal H}_{CQD}, \rho(t)]
+{\cal D}[{\cal T}+{\cal X} n_1]\rho(t)
\label{masterEq}
\\
&=&-\frac{i}{\hbar}[{\cal H}_{CQD} -i\hbar({\cal F}^{*}{\cal X} 
-{\cal F}{\cal X}^{*})n_{1}/2, \rho(t)]
+{\cal D}[{\cal X} n_1+ {\cal T}+{\cal F}]\rho(t),
\label{masterEq1}
\\
&\equiv&{\cal L} \rho(t),
\label{Liouvillian}
\end{eqnarray}
\end{mathletters}
where ${\cal D}$ is defined in Eq.\ (\ref{defcalD}).
Here ${\cal F}$ is an arbitrary complex number
\cite{Wiseman94,Wiseman96}, while
we are using ${\cal T}$ and ${\cal X}$
to represent  respectively the quantities ${\cal T}_+$ and ${\cal X}_+$
in Eq.\ (\ref{rate}) evaluated at zero temperature.
%In fact, ${\cal T}$ and ${\cal X}$ in Eq.\ (\ref{masterEq})
%have the same values as
%${\cal T}$ and ${\cal X}$ defined in the last term of Eq.\
%(\ref{rateEqc}).

Requiring that the ensemble average of the conditioned density matrix
$E[\rho_c(t+dt)]=\rho(t+dt)$ satisfies the unconditional master
equation (\ref{masterEquation})
leads to
\begin{equation}
\tilde\rho_{0c}(t+dt) + \tilde\rho_{1c}(t+dt) = (1+dt {\cal L})
\rho_c(t).
\label{measuretilderho}
\end{equation}
Here we have explicitly used the stochastic It\^{o} calculus
\cite{Gardiner85,Oksendal92} for
the definition of time derivatives as
$\dot{\rho}(t)
=\lim_{dt\rightarrow 0}[{\rho(t+dt)-\rho(t)}]/{dt}$.
This is in contrast to the definition,
$\dot{\rho}(t)
=\lim_{dt\rightarrow 0}[{\rho(t+dt/2)-\rho(t-dt/2)}]/{dt}$,
used in another stochastic calculus, the Stratonovich
calculus \cite{Gardiner85,Oksendal92}.
Recall that
Eq.\ (\ref{dN1}) indicates that $E[dN_c(t)]/dt$ equals to
the average electron tunneling rate through the PC barrier.
From Eq.\ (\ref{rate}), the electron tunneling rates
are $D = |{\cal T}|^{2}$ when $n_{1}=0$ and $D'=|{\cal T}+{\cal
X}|^{2}$ when $n_{1}=1$. From Eq.\ (\ref{dN1})
we thus have the correspondence
\begin{equation} 
{\rm Tr} [M_1(dt)\rho_c(t)M_{1}^{\dagger}(dt)]
= {\rm Tr} \{\rho_{c}(t)[{\cal T}^{*}+n_{1}\chi^{*}][{\cal
T}+n_{1}\chi]\}dt.
\label{col}
\end{equation}
Also, for Eq.~(\ref{measuretilderho}) to reproduce
the master equation (\ref{masterEq1}) we must 
have \cite{Wiseman94,Wiseman96}
\begin{equation}
M_{1}(dt) = \sqrt{dt}({\cal X}n_{1}+{\cal T}+{\cal F})
\end{equation}
for some arbitrary complex number ${\cal F}$. By inspection
of Eq.~(\ref{col}) we must have ${\cal F}=0$, so that
\begin{equation}
\tilde\rho_{1c}(t+dt)={\cal J}[{\cal T}+{\cal X} n_1]\rho_c(t){dt}.
\label{measuretilderho1}
\end{equation}
Substituting Eq.\ (\ref{measuretilderho1}) into (\ref{dN1}) yields 
\begin{equation}
 E[dN_c(t)]={\rm Tr}[\tilde{\rho}_{1c} (t+dt)]
 =[D+(D'-D)\langle n_1\rangle_c(t)] dt,
\label{dNav}
\end{equation} 
where $\langle n_1\rangle_c(t)={\rm Tr}[n_1\rho_c(t)]$.
The remaining part, except the jump of Eq.\ (\ref{measuretilderho1}),
on the right hand side of Eq.\ (\ref{measuretilderho})
in time $dt$, corresponds to the effect of a
measurement giving a null result on $\rho_c(t)$:
\begin{equation}
\tilde\rho_{0c}(t+dt) =\rho_c(t)
-dt\left\{ {\cal A}[{\cal T}+{\cal X} n_1]\rho_c(t)
-\frac{i}{\hbar}[{\cal H}_{CQD},\rho_c(t)] \right\},
\label{measuretilderho0}
\end{equation}
where ${\cal A}$ is defined in Eq.\ (\ref{defcalA}).
The corresponding measurement operator is
\begin{equation}
M_0(dt) = 1-dt [(i/\hbar){\cal H}_{CQD}
+(1/2)({\cal T}^*+{\cal X}^*n_1)({\cal T}+{\cal X}\,n_1)].
\label{M2}
\end{equation}

Finally, substituting Eqs.\ (\ref{measuretilderho1}),
(\ref{measuretilderho0}),  (\ref{Trtilderho0}) and (\ref{dNav}) into
Eq.\ (\ref{rhoc}), expanding and keeping the terms of first order in
$dt$, we obtain the stochastic master equation, conditioned on the
observed event in time $dt$:
\begin{equation}
d\rho_c(t)
=dN_c(t)\left [\frac{{\cal J}[{\cal T}+{\cal X} n_1]}
{{\cal P}_{1c}(t)}
-1\right ]\rho_c(t)
+\, dt \left\{
-{\cal A}[{\cal T}+{\cal X} n_1]\rho_c(t)
+{\cal P}_{1c}(t) \rho_c(t)
+\frac{i}{\hbar}[{\cal H}_{CQD},\rho_c(t)] \right\},
\label{condmasterEq}
\end{equation}
where
\begin{equation}
{\cal P}_{1c}(t)= D+(D'-D)\langle n_1\rangle_c(t).
\end{equation}
Note that $dN_c(t)$, from Eq.\ ({\ref{dNav}), is of order $dt$. Hence
terms proportional to $dN_c(t)dt$ are ignored in Eq.\ (\ref{condmasterEq}).
Again averaging this equation over the observed stochastic process by
setting $E[dN_c(t)]$ equal to its expected value, Eq.\ (\ref{dNav}),
gives the unconditional, deterministic
master equation (\ref{masterEq}).
Eq.\ (\ref{condmasterEq}) is one of the main results in this paper.

So far we have assumed perfect detection or efficient measurement.
In this case, the stochastic master equation for the conditioned
density matrix operator (\ref{condmasterEq}) is equivalent to the
following stochastic Sch\"{o}dinger equation (SSE) for the conditioned
state vector:
%\begin{eqnarray}
\begin{equation}
d|\psi_c(t)\rangle
%& = &
=\left[dN_c(t)
\left(\frac{{\cal T}+{\cal X} n_1}
{\sqrt{{\cal P}_{1c}(t)}}
-1\right)
%\right.
%\nonumber \\
%& & \mbox{}
%\left.
-\, dt \left(\frac{i}{\hbar}{\cal H}_{CQD}
+\frac{({\cal T}^*+{\cal X}^*n_1)({\cal T}+{\cal X}\,n_1)}{2}
-\frac{{\cal P}_{1c}(t)}{2}\right)\right]|\psi_c(t)\rangle.
\label{jumpSSE}
\end{equation}
%\end{eqnarray}
This equivalence can be easily verified using the stochastic
It\^{o} calculus \cite{Gardiner85,Oksendal92}
\begin{eqnarray}
d\rho_c(t)&=&d(|\psi_c(t)\rangle\langle\psi_c(t)|)
\nonumber \\
&=& (d|\psi_c(t)\rangle)\, \langle\psi_c(t)|
+ |\psi_c(t)\rangle \,d\langle\psi_c(t)|
+ (d|\psi_c(t)\rangle)\, (d\langle\psi_c(t)|),
\label{Itopsi}
\end{eqnarray}
and keeping terms up to order $dt$.
Since the evolution of the system can be described by a ket state
vector, it is obvious that an efficient measurement or
perfect detection preserves state purity
if the initial state is a pure state.
In this description of the SSE, the quantum average is now
defined, for example, as
$\langle n_1\rangle_c(t)= \langle\psi_c(t)|n_1 |\psi_c(t)\rangle$.
The unconditional density matrix operator is equivalent to the
ensemble average
of {\em quantum trajectories} generated by the SSE,
$\rho(t)=E[|\psi_c(t)\rangle\langle\psi_c(t)|]$, provided that the
initial density operator can be written as
$\rho(0)=|\psi_c(0)\rangle\langle\psi_c(0)|$.

The interpretation \cite{Wiseman93} for the measured system state
conditioned on the measurement, in terms of
gain and loss of information, can be summarized and understood as
follows. In order for the system to be continuously described by a state
vector (rather than a general density matrix), it is necessary
(and sufficient) to have maximal knowledge of its change of state.
This requires perfect detection or efficient measurement, which
recovers and contains all the information lost from the system
to the reservoirs.
If the detection is not perfect and some information about the
system is {\em untraceable}, the evolution of the system can no longer
be described by a pure state vector. For the extreme case of zero
efficiency detection, the information
(measurement results at the detector) carried
away from the system to the reservoirs is (are) completely
ignored, so that the stochastic master equation
(\ref{condmasterEq}) after being averaged over all possible
measurement records
reduces to the unconditional, deterministic master equation
(\ref{masterEq}), leading to decoherence for the system.
This interpretation highlights
the fact that a density matrix operator description of a quantum state
is only necessary when information
is lost irretrievably.
The purity-preserving, conditional state evolution for a pure initial state,
and gradual purification for a non-pure initial state
have been discussed in
Refs.\ \cite{Korotkov99,Korotkov99b,Korotkov00,Korotkov00b}
in the quantum diffusive limit.

%%%%%%%%%%%%%%%%%%%%%%%%%%%%%%%%%%%%%%%%%%%%%%%%%%%%%%%%%%
\section{quantum-diffusive, conditional master equation}
\label{sec:diffusive}
%%%%%%%%%%%%%%%%%%%%%%%%%%%%%%%%%%%%%%%%%%%%%%%%%%%%%%%%%

In this section, we extend the results obtained in previous section and
derive the conditional master equation
when the average electron tunneling current is very large
compared to the extra change of the tunneling current due to the
presence of the electron in the dot closer to the PC.
This limit is studied and
called a ``weakly coupling or responding detector'' limit in Refs.\
\cite{Korotkov99,Korotkov00}.
Here, on the other hand,  we will refer to this
case as quantum diffusion in contrast to the case of quantum jumps.
In the quantum-diffusive limit,
many electrons, ($N>[(D'+D)/(D'-D)]^2\gg 1$), pass
through the PC before one can distinguish which dot is occupied.
In addition, individual electrons tunneling through the PC are ignored
and  time averaging  of the currents is performed.
This allows electron counts, or accumulated electron number, to be
considered as a continuous variable satisfying a
Gaussian white noise distribution.
In Refs.\ \cite{Korotkov99,Korotkov00} a set of
Langevin equations for the random evolution of the CQD system density
matrix elements conditioned on the detector output was presented, based
only on basic physical reasoning. In this section,
we show explicitly, under the quantum-diffusive limit, that our
microscopic approach reproduces \cite{Korotkov-derivation} the rate equations
in Refs.\ \cite{Korotkov99,Korotkov00}.

In quantum optics, a measurement scheme known as  homodyne detection
\cite{Carmichael93,Wiseman93b,Wiseman94} is closely related to the
measurement of the CQD system by a weakly responding PC detector. In
both cases, there is a large parameter to allow the
photocurrent or electron current to be approximated
by a continuous function of time.
We will follow closely the derivation of a smooth master equation
for homodyne detection given in Ref.\
\cite{Wiseman94} (sketched first by Carmichael \cite{Carmichael93})
for the CQD system.

There are two ideal parameters ${\cal T}$ and ${\cal X}$
for the CQD system.
In the quantum-diffusive limit, we
assume $|{\cal T}| \gg |{\cal X}|$ which is consistent with the assumption,
$(D+D')\gg(D'-D)$,  made in Refs.\ \cite{Korotkov99,Korotkov00}
for the weakly coupling or weakly responding PC detector.
Consider the evolution of the system over the short-time interval
$[t,t+\delta t)$. We relate the three parameters, ${\cal X}$, ${\cal T}$
and $\delta t$, in our problem as $|{\cal X}|^2 \,\delta t \sim
\epsilon^{3/2}$, where $\epsilon=(|{\cal X}|/|{\cal T}|)\ll 1$. This scaling
is chosen so that in time $\delta t$, the number of detections (electron
counts) with dot 1 being unoccupied scales as $\delta N \sim
|{\cal T}|^2 \delta t \sim \epsilon^{-1/2} \gg 1$. However, the extra
change in electron number detections due to the presence of the electron in
dot 1 scales as $|{\cal X}|^2 \delta t \sim \epsilon^{3/2} \ll
1$. To be more specific, the average number of detections, following
Eq.\ (\ref{dNav}), up to order of $\epsilon^{1/2}$ is
\begin{equation}
E[\delta N(t)]=|{\cal T}|^2 \delta t
[1+2\epsilon \, \cos\theta\,\langle n_1\rangle_c(t)],
\label{mean}
\end{equation}
where $\theta$ is the relative phase between ${\cal X}$ and ${\cal T}$.
The variance in $\delta N$ will be dominated by the Poisson statistics
of the current $eD=e|{\cal T}|^2$ in time $\delta t$.  Since the number of
counts in time $\delta t$ is very large, the statistics will be approximately
Gaussian. Indeed, it has been shown \cite{Wiseman93b} that the statistics
of $\delta N$ are consistent with that of a Gaussian random variable
of mean given by Eq.\ (\ref{mean}) and the variance up to order of
$\epsilon^{-1/2}$ is $\sigma^2_N=|{\cal T}|^2 \delta t$. The fluctuations
$\sigma^2_N$ is necessarily as large as expressed here in order for
the statistics of $\delta N$ to be consistent with Gaussian
statistics. Thus, $\delta N$ can be approximately written as a
continuous Gaussian random variable \cite{Gardiner85,Oksendal92}:
\begin{equation}
\delta N(t)=\{|{\cal T}|^2 [1+2\epsilon\, \cos\theta\,\langle n_1\rangle_c(t)]
+|{\cal T}| \xi(t)\} \delta t,
\label{deltaN}
\end{equation}
where $\xi(t)$ is a Gaussian white noise characterized by
\begin{equation}
E[\xi(t)]=0, \quad E[\xi(t)\xi(t')]=\delta(t-t').
\label{xi}
\end{equation}
Here $E$ denotes an ensemble average and $\delta(t-t')$ is a delta function.
In stochastic calculus \cite{Gardiner85,Oksendal92},
$\xi(t)dt=dW(t)$ is known as the infinitesimal
Wiener increment. In Eq.\ (\ref{deltaN}), the accuracy in each term is
only as great as the highest order expression in $\epsilon^{1/2}$.
But it is sufficient for the discussions below.

Although the conditional master equation (\ref{condmasterEq})
requires that $dN_c(t)$ to be a point process, it is possible, in the
quantum-diffusive limit, to simply replace $dN_c(t)$ by the continuous
random variable $\delta N_c(t)$, Eq.\ (\ref{deltaN}). This is because
each jump is infinitesimal, so the effect of many jumps is
approximately equal to the effect of one jump scaled by the number of
jumps. This can be justified more rigorously as in Ref.\
\cite{Wiseman93b}. Finally, expanding Eq.\ (\ref{condmasterEq}) in
power of $\epsilon$, substituting $dN_c(t)\rightarrow\delta N_c(t)$,
keeping only the terms up to the order $\epsilon^{3/2}$, and letting
$\delta t \rightarrow dt$, we obtain the conditional master equation
\begin{eqnarray}
%\begin{equation}
\dot{\rho}_c(t)
&=&
%=
-\frac{i}{\hbar}[{\cal H}_{CQD}, \rho_c(t)]
+{\cal D}[{\cal T}+{\cal X}n_1]\rho_c(t)
\nonumber \\
&&
+\xi(t)\frac{1}{|{\cal T}|}
\left[{\cal T}^* {\cal X}\,n_1\rho_c(t)+ {\cal X}^* {\cal T}\rho_c(t)n_1
-2\,{\rm Re}({\cal T}^* {\cal X})\langle n_1\rangle_c(t)\rho_c(t)\right].
\label{diffusivemasterEq}
%\end{equation}
\end{eqnarray}
%where the superoperator ${\cal D}$ is defined in Eq.\ (\ref{defcalD})
%Re$({\cal T}^* {\cal X})$ stands for taking
%the real part of ${\cal T}^* {\cal X}$.
Thus the quantum-jump evolution of Eq.\ (\ref{condmasterEq}) has been
replaced by quantum-diffusive evolution, Eq.\ (\ref{diffusivemasterEq}).
Following the same reasoning in obtaining the SSE (\ref{jumpSSE})
for the case of quantum-jump process,
we find the quantum-diffusive, conditional master equation
(\ref{diffusivemasterEq}) is equivalent to the following diffusive SSE:
\begin{eqnarray}
%\begin{equation}
d|\psi_c(t)\rangle
& = &
%=
\left[ dt \left(-\frac{i}{\hbar}{\cal H}_{CQD}
-\frac{|{\cal X}|^2}{2}[n_1
-2 n_1 \langle n_1\rangle_c(t)
+ \langle n_1\rangle_c^2(t)]-i\,{\rm Im}({\cal T}^* {\cal X})n_1] \right)
\right.
\nonumber \\
&& +\left.
%+
\xi(t) dt \frac{1}{|{\cal T}|}
\left\{{\cal T}^*{\cal X} n_1-{\cal X}^*{\cal T} \,
\langle n_1\rangle_c(t) \right\}
\right] \, |\psi_c(t)\rangle.
\label{diffusiveSSE}
%\end{equation}
\end{eqnarray}
This equivalence can be verified using Eq.\ (\ref{Itopsi})
and keeping terms up to oder $dt$. Note however in
this case \cite{Gardiner85,Oksendal92} that
terms of order $\xi(t)dt$ are to be
regarded as the same order as $dt$, but $[\xi(t)dt]^2=[dW(t)]^2=dt$.

Our conditional master equation by its
derivation is formulated in terms of It\^{o} calculus, while the
stochastic rate equations in Refs. \cite{Korotkov99,Korotkov00} are
written in a Stratonovich calculus form \cite{Gardiner85,Oksendal92}.
In contrast to the Stratonovich
form of the rate equations, it is easy to see that the ensemble
average evolution of our conditional master equation
(\ref{diffusivemasterEq}) reproduces the unconditional master equation
(\ref{masterEq}) by simply eliminating the white noise term using
Eq.\ (\ref{xi}).
To show that our quantum-diffusive, conditional stochastic
master equation (\ref{diffusivemasterEq})
reproduces the non-linear Langevin rate
equations obtained semi-phenomenologically in Refs.\
\cite{Korotkov99,Korotkov00}, we
evaluate Eq.\ (\ref{diffusivemasterEq})
in the same basis as for Eq.\ (\ref{rateEq}) and
obtain:
\begin{mathletters}
\label{diffrateEq}
\begin{eqnarray}
\dot\rho_{aa}(t)& = &i\Omega[\rho_{ab}(t)-\rho_{ba}(t)]
-\sqrt{8\Gamma_d} \, \rho_{aa}(t) \rho_{bb}(t)\xi(t)\;,
\label{diffrateEqa}\\
\dot\rho_{ab}(t)& = & i\varepsilon \, \rho_{ab}(t)+i\Omega[\rho_{aa}(t)
-\rho_{bb}(t)]-\Gamma_d \, \rho_{ab}(t)
+\sqrt{2\Gamma_d}\, \rho_{ab}(t)[\rho_{aa}(t)-\rho_{bb}(t)]\xi(t)\;,
\label{diffrateEqc}
\end{eqnarray}
\end{mathletters}
In obtaining Eq.\ (\ref{diffrateEq}),
we have made the assumption of real tunneling amplitudes as in
Refs.\ \cite{Gurvitz97,Korotkov99,Korotkov00}
in order to be able to compare the results directly.
We have also set ${\cal X}=-\sqrt{2 \Gamma_d}$.
Again, the ensemble average of Eq.\ (\ref{diffrateEq})
by eliminating the white noise terms
reduces to Eq.\ (\ref{rateEq}).
To further demonstrate the equivalence, we translate the stochastic rate
equations of
Refs.\ \cite{Korotkov99,Korotkov00} into It\^{o} formalism
and compare them to Eq.\ (\ref{diffrateEq}).
This is carried out in Appendix \ref{sec:equivalence}.
Indeed, Eq.\ (\ref{diffrateEq}) is equivalent to the
Langevin rate equations in
Refs.\ \cite{Korotkov99,Korotkov00} for the ``ideal detector''.

%%%%%%%%%%%%%%%%%%%%%%%%%%%%%%%%%%%%%%%%%%%%%%%
\section{Analytical results for conditional dynamics}
\label{sec:analytic}
%%%%%%%%%%%%%%%%%%%%%%%%%%%%%%%%%%%%%%%%%%%%%%

To analyze the dynamics of a two-state system, such as the CQD system
considered here, one can represent the system density matrix elements
in terms of Bloch sphere variables. The Bloch sphere representation is
equivalent to that of the rate equations. However, some
physical insights into the dynamics of the system can sometimes be
 more easily visualized in this representation. Denoting the averages of the
operators $\sigma_x$, $\sigma_y$, $\sigma_z$ by $x$, $y$, $z$
respectively, the density matrix operator for the CQD system can be
expressed in terms of the Bloch sphere
vector $(x,y,z)$ as:
\begin{mathletters}
\label{Blochsphere}
\begin{eqnarray}
\rho(t)&=&[I+x(t)\sigma_x
+y(t)\sigma_y+z(t)\sigma_z]/2
\label{Blochrho}
\\
&=&\frac{1}{2}\left (\begin{array}{cc}
1+z(t)&x(t)-iy(t)\\
x(t)+iy(t)& 1-z(t)\end{array}\right ),
\label{bloch}
\end{eqnarray}
\end{mathletters}
where the operator $I$, $\sigma_i$, are defined using the fermion
operators for the two dots:
\begin{mathletters}
\label{sigma}
\begin{eqnarray}
I&=& c^\dagger_{2}c_{2}+c^\dagger_{1}c_{1},\\
\sigma_x & = & c^\dagger_{2}c_{1}+c^\dagger_{1}c_{2}, \\
\sigma_y & = & -ic^\dagger_{2}c_{1}+ic^\dagger_{1}c_{2}, \\
\sigma_z & = & c^\dagger_{2}c_{2}-c^\dagger_{1}c_{1}.
\end{eqnarray}
\end{mathletters}
It is easy to see that ${\rm Tr}\rho(t)=1$, $I$ is a unit operator, and
$\sigma_i$ defined above satisfies the
properties of Pauli matrices.
In this representation, the variable $z(t)$ represents the
population difference between the two dots. Especially, $z(t)=1$ and
$z(t)=-1$ indicate that the electron is localized in dot 2 and
dot 1 respectively. The value $z(t)=0$ corresponds to an equal
probability for the electron to be in each dot.

The master equations (\ref{masterEq}), (\ref{diffusivemasterEq})
and(\ref{condmasterEq}),
can be written as a set of coupled stochastic
differential equations in terms of the Bloch sphere variables.
For simplicity, in this section we assume that
the tunneling amplitudes are real.
By substituting Eq.\ (\ref{Blochrho}) into Eq.\ (\ref{masterEq}),
and collecting and equating the coefficients in front of $\sigma_x$,
$\sigma_y$, $\sigma_z$ respectively, the unconditional master equation
under the assumption of real tunneling amplitudes
is equivalent to the following equations:
\begin{mathletters}
\label{uncondBloch}
\begin{eqnarray}
\frac{d}{dt}\left (\begin{array}{c}
x(t)\\y(t)
\end{array}\right)&=& \left (\begin{array}{cc}
-\Gamma_d & -\varepsilon\\
\varepsilon & -\Gamma_d
\end{array}\right)
\left (\begin{array}{c}
x(t)\\y(t)
\end{array}\right)
+\left (\begin{array}{c}
0\\-2\Omega \, z(t)
\end{array}\right),
\label{uncondxy}
\\
\frac{dz(t)}{dt} &=&2\Omega \, y(t).
\label{uncondz}
\end{eqnarray}
\end{mathletters}
Similarly for the quantum-diffusive, conditional
master equation (\ref{diffusivemasterEq}), we obtain
\begin{mathletters}
\label{diffBlochEq}
\begin{eqnarray}
\frac{dx_c(t)}{dt} &=&-\varepsilon \, y_c(t)-\Gamma_d \, x_c(t)
-\sqrt{2\Gamma_d} \, z_c(t) x_c(t) \xi(t), \\
\frac{dy_c(t)}{dt} &=&\varepsilon x_c(t) -2\Omega \, z_c(t)
-\Gamma_d \, y_c(t) -\sqrt{2\Gamma_d}\, z_c(t) y_c(t) \xi(t),  \\
\frac{dz_c(t)}{dt} &=&2\Omega \, y_c(t)
+\sqrt{2\Gamma_d}\left[1-z^2_c(t)\right] \xi(t).
\label{diffz}
\end{eqnarray}
\end{mathletters}
Again the $c$-subscript is to emphasize that these variables refer to
the conditional state. It is trivial to see that Eq.\
(\ref{diffBlochEq})
averaged over the white noise
reduces to Eq.\ (\ref{uncondBloch}), provided that
$E[x_c(t)]=x(t)$ as well as similar replacements are performed
for $y_c(t)$ and $z_c(t)$.
The analogous calculation can be carried out for the quantum-jump,
conditional master equation (\ref{condmasterEq}). We obtain
\begin{mathletters}
\label{condBlochEq}
\begin{eqnarray}
dx_c(t)&=&dt \left(-\varepsilon \, y_c(t)-\frac{(D'-D)}{2} z_c(t) x_c(t)\right)
-dN_c(t)\left(
x_c(t)\frac{2\Gamma_d-(D'-D)z_c(t)}{2D+(D'-D)[1-z_c(t)]}\right)\, ,
\\
dy_c(t) &=&dt\left( \varepsilon \, x_c(t)
-2\Omega \, z_c(t)-\frac{(D'-D)}{2} z_c(t) y_c(t)\right)
-dN_c(t)\left(
y_c(t)\frac{2\Gamma_d-(D'-D)z_c(t)}{2D+(D'-D)[1-z_c(t)]}\right) \, ,
\\
dz_c(t) &=&dt \left(2\Omega \, y_c(t)dt
+\frac{(D'-D)}{2}\left[1-z_c^2(t)\right]\right)
-dN_c(t)\left(\frac{(D'-D)[1-z^2_c(t)]}{2D+(D'-D)[1-z_c(t)]}\right) \, .
\label{condz}
\end{eqnarray}
\end{mathletters}
As expected, by using Eq.\ (\ref{dNav}), the ensemble average
of Eq.\ (\ref{condBlochEq}) also reduces to the unconditional
equation (\ref{uncondBloch}).

Next we calculate the localization rate, at which the electron
becomes localized in one of the two dots due to the measurement,
using Eqs.\ (\ref{diffBlochEq}) and (\ref{condBlochEq}).
Obviously, the stochastic, conditional differential equations provide more
information than the unconditional ones do. In the unconditional case
Eq.\ (\ref{uncondBloch}), the
average population difference $z(t)$ between the dots
is a constant of motion ($[dz(t)/dt]=0$) when $\Omega=0$. However, if
the present model indeed describes a measurement of $n_1=c_1^\dagger
c_1$ (in other words the position of the electron in the dots),
then in the absence of tunneling $\Omega=0$,
we would expect to see the conditional state
become localized in one of the two
dots, i.e., either $z=1$ or $z=-1$. Indeed, for $\Omega=0$,
we can see from the conditional equations (\ref{diffz}) and (\ref{condz})
that $z_c(t)=\pm 1$ are fixed points. We can take into account both
fixed points by considering $z_c^2(t)$.
Hence it is sensible to take the
ensemble average $E[z_c^2(t)]$ and find the rate at which this
deterministic quantity approaches one.
applying It\^{o} calculus \cite{Gardiner85,Oksendal92}
to the stochastic variable $z_c^2(t)$,
we have $d(z_c^{2})=2z_cdz_c+dz_cdz_c$. Let us first consider
the case for the quantum-jump equations.
Using Eqs.\ (\ref{condz}) and (\ref{dNav}) and the fact that
$dN_c^2(t)=dN_c(t)$, we find that
\begin{equation}
E[dz_c^2(t)]=E
\left[ \frac{(D'-D)^2 [1-z_c^2(t)]^2}{4D+2(D'-D)[1-z_c(t)]}\right] \,dt.
\label{jumpdz2}
\end{equation}
If the system starts in a state which has an equal probability for the
electron to be in each dot then $z_c(0)=z(0)=0$. In this case, the ensemble
average variable $z(t)$ would remain to be zero since $[dz(t)/dt]=0$
when $\Omega=0$.
However if we average $z_c^2$ over many quantum trajectories
with this initial condition then we find from Eq.\ (\ref{jumpdz2})
that for short times (by setting $z_c(t), z_c^2(t)\approx 0$)
\begin{equation}
E[z_c^2(\delta t)]\approx \frac{(D'-D)^2}{2(D'+D)}\, \delta t
=\gamma^{\rm jump}_{\rm loc} \, \delta t.
\label{jumpEz2}
\end{equation}
That is to say, the system tends toward a definite state
(with $z_c=\pm 1$ so $z_c^2=1$) at an initial rate of
\begin{equation}
\gamma^{\rm jump}_{\rm loc}=\frac{(D'-D)^2}{2(D'+D)}
=\frac{(\sqrt{D'}+\sqrt{D})^2}{(D'+D)} \,\Gamma_d.
\label{jumplocrate}
\end{equation}
Similarly for the case of quantum diffusion, using Eqs.\ (\ref{condz})
and (\ref{xi}) and the fact that $[\xi(t)dt]^2=[dW(t)]^2=dt$,
we find
$E[dz_c^2(t)]=E\left[2\Gamma_d [1-z_c^2(t)]^2 \right] \,dt$.
Applying the same reasoning for obtaining Eq.\ (\ref{jumpEz2}), we find
$E[z_c^2(\delta t)]\approx 2\Gamma_d \, \delta t
=\gamma^{\rm diff}_{\rm loc} \, \delta t$.
This implies that the localization rate in this case
is $\gamma^{\rm diff}_{\rm loc}=2\Gamma_d$.
This is consistent with the result of localization
time, $t_{\rm loc}\sim (1/\gamma^{\rm diff}_{\rm loc})$,
found in Ref.\ \cite{Korotkov99}
in the quantum-diffusive case.
As expected, Eq.\ (\ref{jumplocrate}) in the quantum-diffusive limit,
${\cal T} \gg {\cal X}$ or $(D+D')\gg(D'-D)$, reduces to
$\gamma^{\rm jump}_{\rm loc} \rightarrow 2 \Gamma_d
=\gamma^{\rm diff}_{\rm loc}$.
The rate of localization is a direct indication of
the quality of measurement. It is
necessarily as large as the decoherence rate
since a successful measurement distinguishing
the location of the electron on the two dots
would destroy any coherence between them.

The above localization rates are related to the
signal-to-noise ratio for the measurement and can be
obtained intuitively as follows.
Consider the electron with equal
likelihood in either dot so that $z_c(0)=z(0)=0$.
For the case of quantum diffusion, the electron
tunneling current through the PC obeys Gaussian statistics.
Recall in Sec.\ \ref{sec:diffusive} that the mean of
the probability distribution of the  number of electron detections
through the PC is given
by Eq.\ (\ref{mean}) and its variance takes the form
$\sigma^2_N={\cal T}^2 \delta t$ in time $\delta t$.
If the electron is in dot 1, then
the rate of electrons passing through the PC is
${\cal T}^2+2{\cal T}{\cal X}$;
if it is in dot 2, then the rate is ${\cal T}^2$.
One may define the width of the probability distribution as
the distance from the mean when the distribution
falls to $e^{-1}$ of its maximum value.
For a Gaussian distribution, the square of the
width is twice the variance.
The above two probability distributions will begin to be
distinguishable when the difference in
the means of the two distributions
is of order the square root of the sum of twice the
variances (square of the widths) at time $\tau$.
That is, when
\begin{equation}
[({\cal T}^2+2{\cal T}{\cal X})\tau -{\cal T}^2\tau]
\sim \sqrt{2{\cal T}^2\tau + 2{\cal T}^2\tau}.
\label{difftloc}
\end{equation}
Solving this for $\tau$ gives a characteristic rate:
$\tau^{-1} \sim  {\cal X}^2=2 \Gamma_d$.
This is just the $\gamma^{\rm diff}_{\rm loc}$ discussed above.
For the case of quantum jumps, the statistics of
the electron counts through
the PC can be approximated by
Poisson statistics.
For a Poisson process at rate ${\cal R}$,
the probability for $N$ events to occur
in time $t$ is
\begin{equation}
p(N;t)=\frac{({\cal R} t)^N}{N!}e^{-{\cal R} t}.
\end{equation}
The mean and variance of this distribution are equal and given by
$E[N] = {\rm Var}(N) = {\cal R}t$.
In the quantum-jump case from Eq.\ (\ref{dNav}),
if the electron is in dot 1 then
the rate of electrons passing through the PC is $D'$.
If the electron is in dot 2, then the rate is just $D$.
Requiring the difference in means of the two probability
distributions, $p(N,\tau)$, being of order the square root
of the sum of twice
the variances at time $\tau$ yields:
\begin{equation}
(D'\tau -D\tau) \sim \sqrt{2 D'\tau + 2 D\tau}.
\label{jumptloc}
\end{equation}
Solving this for $\tau^{-1}$ yields a characteristic rate
which is the same as $\gamma^{\rm jump}_{\rm loc}$
defined in Eq.\ (\ref{jumplocrate}).

A similar conclusion is reached in
Refs.\ \cite{Shnirman98,Makhlin98,Makhlin00}.
The measurement time, $t_{\rm ms}$, in
Refs.\ \cite{Shnirman98,Makhlin98,Makhlin00}
is roughly the inverse of the localization rate
given here.
However, there the condition for being able to distinguish
the two probability distribution is slightly
different from the condition discussed here.
The measurement time \cite{Shnirman98,Makhlin98,Makhlin00} is
denoted as the time at which the
separation in the means of the two distributions
is larger than the sum of the widths, i.e.,
the sum of the square roots of twice the individual variance
rather than the square root of the sum of twice
the variance.
If this condition is applied here, instead of
Eqs.\ (\ref{difftloc}) and (\ref{jumptloc}), we have
\begin{equation}
[({\cal T}^2+2{\cal T}{\cal X}) t_{\rm ms} -{\cal T}^2 t_{\rm ms}]
\geq \sqrt{2{\cal T}^2 t_{\rm ms}}+ \sqrt{2{\cal T}^2 t_{\rm ms} }
\label{difftms}
\end{equation}
for the quantum-diffusive case, and
\begin{equation}
(D' \, t_{\rm ms} -D \, t_{\rm ms} )
\geq \sqrt{2 D' \, t_{\rm ms} }+ \sqrt{2 D \, t_{\rm ms} }
\label{jumptms}
\end{equation}
for the quantum-jump case.
We find from Eqs.\ (\ref{difftms}) and (\ref{jumptms}) that the inverse
of the measurement
time $t_{\rm ms}$ is the same for both quantum-diffusive and
quantum-jump cases, and is equal to the decoherence rate:
\begin{equation}
t^{-1}_{\rm ms}=\frac{{\cal X}^2}{2}
=\frac{(\sqrt{D'}-\sqrt{D})^2}{2}
=\Gamma_d=\tau_d^{-1},
\end{equation}
where $\tau_d=(1/\Gamma_d)$
is the decoherence time.
This is in agreement with the result in
Refs.\ \cite{Makhlin98,Makhlin00}.
Our condition shows, on the other hand, the different localization
rates for the quantum-jump and quantum-diffusive cases.
This is consistent with the initial rates obtained
from the ensemble average of Bloch variable, $ E[z_c^2(\delta t)]$.

There is another time scale denoted as mixing time, $t_{\rm mix}$,
discussed in Refs.\ \cite{Shnirman98,Makhlin98,Makhlin00}.
It is the time after which the information about
the initial quantum state of the CQDs
is lost due to the measurement-induced transition.
This transition arises because of the non-zero
coupling $\Omega$ term in the CQD Hamiltonian,
which does not commute with the occupation number operator of dot 1
(the measured quantity) and thus mixes
the two possible states of the CQD system.
Below we estimate the mixing time using the differential equations
for the Bloch variables.
It is expected that effective and successful quantum measurements
require $t_{\rm mix} \gg t_{\rm ms} \sim t_{\rm loc} \sim \tau_d$.
In other words, the readout should be achieved
long before the information about
the measured initial quantum state is lost.
In terms of different characteristic rates, we have,
in this case, the relation:
$\Gamma_d \sim \gamma_{\rm loc} \sim t^{-1}_{\rm ms} \gg \gamma_{\rm mix}$,
where $\gamma_{\rm mix}=(1/t_{\rm mix})$ represents the mixing rate.
For finite $\Omega$,
the rate at which the variables $x(t)$ and $y(t)$ relax can be found
from the real part of the eigenvalues of the matrix
in the first term on the right hand side of Eq.\ (\ref{uncondxy}).
This gives the decay (decoherence) rate $\Gamma_d$ for the
off-diagonal variables, $x(t)$ and $y(t)$.
The variable $z(t)=0$ represents an equal
probability for the electron in the CQDs to be in each dot.
Hence the rate at which the variable $z(t)$ relaxes to zero
corresponds to the mixing rate \cite{Makhlin00}, $\gamma_{\rm mix}$.
Under the assumption of $\Gamma_d \gg \gamma_{\rm mix}$
for effective measurements,
the variables $x(t)$ and $y(t)$ therefore
relax at a rate much faster than
that of the variable $z(t)$.
As a result,
it is valid to substitute the steady-state value of $y(t)$
obtained from Eq.\ (\ref{uncondxy})
into $\dot{z}(t)$ Eq.\ (\ref{uncondz}) to find the mixing rate.
Consequently, we obtain
\begin{equation}
\frac{dz(t)}{dt}
=-\frac{4\, \Omega^2 \, \Gamma_d}{\Gamma_d^2+\varepsilon^2} \, z(t)
=-\gamma_{\rm mix} \, z(t).
\label{mixrate}
\end{equation}
It is easy to see that the mixing rate Eq.\ (\ref{mixrate})
vanishes as $\Omega\rightarrow 0$.
Finally, the self-consistent requirement for the assumption
$\Gamma_d \gg \gamma_{\rm mix}$ yields,
from Eq.\ (\ref{mixrate}), the condition:
$\Omega \ll (\sqrt{\Gamma_d^2+\varepsilon^2}/2)$.
The mixing rate Eq.\ (\ref{mixrate}) is in agreement with
the result found in Ref.\ \cite{Makhlin00} under the
similar required condition \cite{mixingrate}.

%%%%%%%%%%%%%%%%%%%%%%%%%%%%%%%%%%%%%%%%%%%%%
\section{Conclusion}
\label{sec:conclusion}
%%%%%%%%%%%%%%%%%%%%%%%%%%%%%%%%%%%%%%%%%%%%%
We have obtained the unconditional master equation for the CQD system,
taking into account the effect of finite-temperature of the PC
reservoirs under the weak system-environment coupling
and Markovian approximations.
We have also presented a
{\em quantum trajectory} approach
to derive, for both quantum-jump and quantum-diffusive cases, the
zero-temperature conditional master equations.
These conditional master equations describe the evolution of the measured
CQD system, conditioned on a particular realization of the measured current.
We have found in both cases that the dynamics of the CQD system can be
described by the SSEs for its
conditional state vector provided that the information carried away
from the system by the PC reservoirs can be recovered by perfect
 measurement detection. Furthermore,
we have analyzed for both cases the localization rates at which the
electron becomes localized in one of the two dots when $\Omega=0$.
We have shown that the localization time discussed here
is slightly different from
the measurement time defined in
Refs.\ \cite{Shnirman98,Makhlin98,Makhlin00}.
The mixing rate at which the two possible states of
the CQDs become mixed when $\Omega\neq 0$ has been calculated as well
and found in agreement with
the result in Ref.\ \cite{Makhlin00}.

%%%%%%%%%%%%%%%%%%%%%%%%%%%%%%%%%%%%%%%%%%%%%
%\section{Acknowledgment}
%\label{sec:Acknow}
%%%%%%%%%%%%%%%%%%%%%%%%%%%%%%%%%%%%%%%%%%%%%
We thank M.~B\"{u}ttiker and A.~M.~Martin for
bringing our attention to Ref.\ \cite{Buttiker00}.
HSG is grateful to useful discussions with A.~N.~Korotkov,
D.~V.~Averin, K.~K.~Likharev, D.~P.~DiVincenzo, J.~R.~Friedman and
V.~Sverdlov.

%%%%%%%%%%%%%%%%%%%%%%%%%%%%%%%%%%%%%%%%%%%%%%
\appendix
%%%%%%%%%%%%%%%%%%%%%%%%%%%%%%%%%%%%%%%%%%%%%%
%%%%%%%%%%%%%%%%%%%%%%%%%%%%%%%%%%%%%%%%%%%%%%
\section{Equivalence of stochastic rate equations in different calculus}
\label{sec:equivalence}
%%%%%%%%%%%%%%%%%%%%%%%%%%%%%%%%%%%%%%%%%%%%%%
In this Appendix, we translate the stochastic rate equations of
Refs.\ \cite{Korotkov99,Korotkov00}, written in terms of Stratonovich
calculus, into It\^{o} calculus formalism
\cite{Gardiner85,Oksendal92}. Although the translation
was sketched and the result was stated in Ref.\ \cite{Korotkov99}, for
completeness, we fill in the calculation steps using our
notation here.
Eq.\ (11) and (12) of Ref.\ \cite{Korotkov99} in
Stratonovich calculus formalism are rewritten
in terms of our notation as follows:
\begin{eqnarray}
\dot{\rho}_{bb}(t) &=& i\Omega[\rho_{ba}(t)-\rho_{ab}(t)]
 +4\sqrt{\frac{\Gamma_d}{S_I}}\, \rho_{bb}(t)\rho_{aa}(t)
\left(-\sqrt{S_I \Gamma_d} [\rho_{aa}(t)-\rho_{bb}(t)]
+\sqrt{\frac{S_I}{2}} \xi (t)\right),
\label{Bayes1}\\
 \dot{\rho}_{ba}(t)&=& -i\varepsilon\rho_{ba}(t)
+i\Omega[\rho_{bb}(t) -\rho_{aa}(t)]
\nonumber \\
&&  + 2\sqrt{\frac{\Gamma_d}{S_I}}\, \, [ \rho_{aa}(t)-\rho_{bb}(t)]
\left(-\sqrt{S_I \Gamma_d} [\rho_{aa}(t)-\rho_{bb}(t)]
+\sqrt{\frac{S_I}{2}} \xi(t)\right) \, \rho_{ba}(t),
\label{Bayes2}
\end{eqnarray}
where we have substituted the notation used in Ref.\ \cite{Korotkov99}
to $H/\hbar\to\Omega$, $\varepsilon/\hbar\to\varepsilon$, and
expressions for ${\cal R}$ and $\Delta I$
in terms of $\Gamma_d$ and $S_I$ using Eqs.\ (10) and (2) of
Ref.\ \cite{Korotkov99}.
Specifically, we have set
$\Delta I=-2 \sqrt{S_I \Gamma_d}$.
In addition, the white noise $\xi(t)$ in Ref.\ \cite{Korotkov99}
has spectral density
$S_\xi=S_I$, which implies $E[\xi(t)\xi(t')]=(S_I/2)\delta(t-t')$,
different from our definition, Eq.\ (\ref{xi}). Hence, the replacement
$\xi(t)\rightarrow\sqrt{S_I/2} \, \xi(t)$ has been employed.
Moreover, since an
ideal detector is assumed, $\gamma_d$ is set to zero for Eq.\ (12) of
Ref.\ \cite{Korotkov99}.
Note finally that the electron operator indices in the CQD should
be interchanged. For example,
the electron annihilation operator $c_2$
in Ref.\ \cite{Korotkov99} should be
$c_1$ in our notations. As a result, $\rho_{11}(t)$
in Ref.\ \cite{Korotkov99} is rewritten as $\rho_{bb}(t)$,
and $\rho_{12}(t)$ as $\rho_{ba}(t)$ here.

As pointed out in Ref.\ \cite{Korotkov99}, to translate
Eqs.\ (\ref{Bayes1}) and (\ref{Bayes2})
into It\^o formalism, one needs to add the term \cite{Gardiner85,Oksendal92}
$ (F/2)(dF/d\rho_{ij})$ for each rate equation $\dot{\rho}_{ij}(t)$,
where $F$ is the factor before $\xi (t)$ in each equation
respectively.
Note that the factor $S_I/2$ appearing in front of the term
needed to be added for the translation in Ref.\ \cite{Korotkov99}
is set to $1$ here. This is because of the different
definitions of the stochastic white noise variables $\xi(t)$ in both
cases, discussed above. To be more specific, $F= \sqrt{8\Gamma_d}\,
\rho_{bb}(t)\rho_{aa}(t)$ for Eq.\ (\ref{Bayes1}) in our notations. By using
the relation $\rho_{aa}(t)=1- \rho_{bb}(t)$, it is easy to find the
derivative $dF/d\rho_{bb}=\sqrt{8\Gamma_d}\,[\rho_{aa}(t)-\rho_{bb}(t)]$. As
a consequence, the term needed to be added to Eqs.\ (\ref{Bayes1}) is
\begin{equation}
4\Gamma_d\, \rho_{bb}(t)\rho_{aa}(t) [\rho_{aa}(t)-\rho_{bb}(t)],
\end{equation}
which exactly cancels
the first term inside the big parenthesis
in the second term of Eq.\ (\ref{Bayes1}).
Hence the resultant equation for Eq.\ (\ref{Bayes1}) in It\^{o} form
is just Eq.\ (\ref{diffrateEqa}) with an overall minus sign in front
of it ($\dot{\rho}_{bb}(t)=-\dot{\rho}_{aa}(t)$).
As for Eq.\
(\ref{Bayes2}), it is easy to find that
$F= \sqrt{2\Gamma_d}\, [\rho_{aa}(t)-\rho_{bb}(t)]\rho_{ba}(t)$.
In order to carry out the derivative with respect to $\rho_{ba}(t)$,
one needs the expression of Eq.\ (8) of Ref.\ \cite{Korotkov99} to
relate diagonal elements, $\rho_{bb}(t)$ and $\rho_{aa}(t)$,
to $\rho_{ba}(t)$.  We then obtain
\begin{equation}
\frac{dF(t)}{d\rho_{ba}(t)}
=\sqrt{2\Gamma_d}\left[
\frac{2[\rho_{aa}(t)-\rho_{bb}(t)]^2 -1}{\rho_{aa}(t)-\rho_{bb}(t)}\right] \, .
\end{equation}
Thus the terms needed to be added into Eq.\ (\ref{Bayes2}) are
\begin{equation}
2\Gamma_d[\rho_{aa}(t)-\rho_{bb}(t)]^2 \rho_{ba}(t)-\Gamma_d\,  \rho_{ba}(t).
\label{add2}
\end{equation}
The first term in Eq.\ (\ref{add2}) exactly cancels the term
with the square bracket
inside the big parenthesis in the last term of Eq.\ (\ref{Bayes2}).
Therefore the resultant
equation for Eq.\ (\ref{Bayes2}) in It\^{o} form is equal to
the complex conjugate of Eq.\ (\ref{diffrateEqc})
($\dot{\rho}_{ba}(t)=\dot{\rho}_{ab}^*(t)$).
This completes our demonstration of the equivalence.

%%%%%%%%%%%%%%%%%%%%%%%%%%%%%%%%%%%%%%%%%%%%%%%%%%%%%%%%

%%%%%%%%%%%%%%%%%%%%%%%%%%%%%%%%%%%%%%%%%%%%%%%%%
%figures
%%%%%%%%%%%%%%%%%%%%%%%%%%%%%%%%%%%%%%%%%%%%%%%%%
\newpage

\begin{figure}
\centerline{\psfig{file=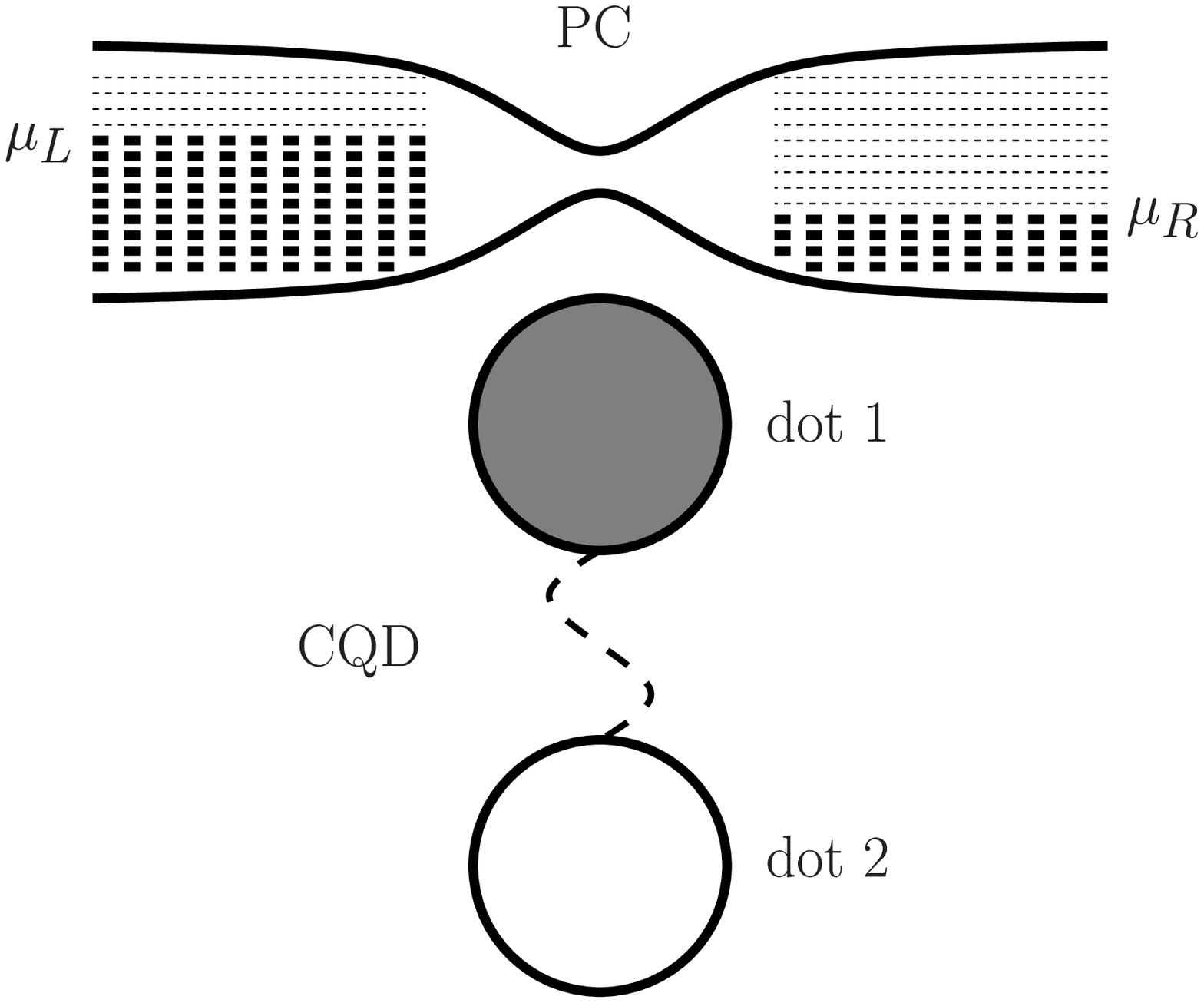,width=\linewidth,angle=0}}
\caption{Schematic representation of two coupled quantum dots (CQD)
when one dot is subjected to a measurement of its electron
occupation number using a low-transparency point contact (PC) or
tunnel junction.
Here $\mu_L$ and $\mu_R$ stand for the
chemical potentials in the left and right
reservoirs respectively.}
\label{fig:PC}
\end{figure}


\begin{references}
%%%%%%%%%%%%%%%%%%%%%%%%%%%%%%%%%%%%%%%%%%%%%%%%%%%%%%%%

\bibitem[*]{goan} E-mail: goan@physics.uq.edu.au.

\bibitem{Datta95}  S.~Datta, {\em Electronic Transport in Mesoscopic Systems},
        (Cambridge University Press, Cambridge, 1995); Y.~Imry, {\it
        Introduction to Mesoscopic Physics},
        (Oxford University Press, New York, 1997).

\bibitem{Schon97} T.~Dittrich, P.~H\"{a}nggi, G.-L.~Ingold, B.~Kramer,
        G.~Sch\"on, and W.~Zwerger, {\it Quantum Transport and
        Dissipation} (Wiley-Vch, Weinheim, 1997).

\bibitem{Buks98} E.~Buks, R.~Schuster, M.~Heiblum, D.~Mahalu,
        and V.~Umansky, Nature {\bf 391}, 871 (1998).

\bibitem{Nakamura99} Y.~Nakamura, Yu.~A Pashkin, and J.~S.~Tsai,
        Nature {\bf 398}, 786 (1999).

\bibitem{Sprinzak99} D.~Sprinzak, E.~Buks, M.~Heiblum, and H.~Shtrikman,
        cond-mat/9907162.


\bibitem{Kane98}  B.~E.~Kane, Nature, {\bf 393}, 133 (1998).
              V.~Privman, I.~D.~Vagner, and G.~Kventsel,
              Phys. Lett. A 239, 141 (1998).


\bibitem{Loss98}  D.~Loss and D.~P.~DiVincenzo, Phys. Rev. A 57, 120 (1998).
                  A.~Imamoglu, D.~D.~Awschalom, G.~Burkard, D.~P.~DiVincenzo,
                  B.~D.~Loss, M.~Sherwin, and A.~Small, Phys. Rev. Lett. 83,
                  4204 (1999).

\bibitem{Schoen97}  A.~Shnirman, G.~Sch\"on, and Z.~Hermon,
                      Phys. Rev. Lett. 79, 2371 (1997);
                    D.~V.~Averin, Solid State Commun. 105, 659 (1998).

\bibitem{Bonadeo98}  N.~H.~Bonadeo, J.~Erland, D.~Gammon, D.~Park,
                     D.~S.~Katzer, and D.~G.~Steel, Science 282, 1473 (1998).

\bibitem{Wheeler83} {\it Quantum Theory of Measurement}, ed. by
        J.~A.~Wheeler and W.~H.~Zurek (Princeton Univ. Press,
        1983).

\bibitem{Braginsky92} V.~B.~Braginsky and F.~Ya.~Khalili,
        {\it Quantum measurement} (Cambridge Univ. Press, 1992).


\bibitem{Aleiner97} I.~L.~Aleiner, N.~S.~Wingreen, and Y.~Meir,
        Phys. Rev. Lett. {\bf 79}, 3740 (1997).

\bibitem{Levinson97} Y.~Levinson, Europhys. Lett. {\bf 39}, 299 (1997).

\bibitem{Stodolsky98} L.~Stodolsky, Phys. Lett. B {\bf 459}, 193
                      (1999); quant-ph/9805081.

\bibitem{Buttiker00} M.~B\"{u}ttiker and A.~M.~Martin, Phys. Rev. B
        {\bf 61}, 2737 (2000).

\bibitem{Gurvitz97} S.~A.~Gurvitz, Phys. Rev. B {\bf 56}, 15215
        (1997).
\bibitem{Gurvitz98} S.~A.~Gurvitz, quant-ph/9808058.

\bibitem{Korotkov99} A.~N.~Korotkov, Phys. Rev. B {\bf 60}, 5737
         (1999); quant-ph/9808026.

\bibitem{Korotkov99b} A.~N.~Korotkov, in {\it Proceedings of LT'22}
        (Helsinki, 1999) and cond-mat/9906439.

\bibitem{Korotkov00} A.~N.~Korotkov, cond-mat/0003225.

\bibitem{Makhlin98} Y.~Makhlin, G.~Sch\"on, and A.~Shnirman,
        cond-mat/9811029.


\bibitem{Makhlin00} Y.~Makhlin, G.~Sch\"on, and A.~Shnirman,
        cond-mat/0001423.

\bibitem{Korotkov00b} A.~N.~Korotkov, cond-mat/0008003.

\bibitem{Korotkov00c} A.~N.~Korotkov, cond-mat/0008461.

\bibitem{Hackenbroich98} G.~Hackenbroich, B.~Rosenow, and H.~A.
Weidenm\"{u}ller, Phys. Rev. Lett. {\bf 81}, 5896 (1998).

\bibitem{Averin00}D.~V.~Averin, cond-mat/0004364; A.~N.~Korotkov and
D.~V.~Averin, cond-mat/0002203.

\bibitem{Shnirman98} A.~Shnirman and G.~Sch\"on, Phys. Rev. B {\bf 57},
        15400 (1998).

\bibitem{Wiseman00} H.~M.~Wiseman, D.~W.~Utami, H.~B.~Sun,
         G.~J.~Milburn, B.~E.~Kane, A.~Dzurak, and R.~G.~Clark,
         cond-mat/0002279.

\bibitem{Averin00b}D.~V.~Averin, cond-mat/0008114; cond-mat/0010052.

\bibitem{Brink00}A.~Maassen van den Brink, cond-mat/0009163.

\bibitem{Carmichael93} H.~J.~Carmichael, {\it An open system approach
        to quantum optics}, Lecture notes in physics (Springer, Berlin,
        1993).
\bibitem{Gardiner91}  C.~W.~Gardiner, {\em Quantum Noise}
        (Springer, Berlin, 1991).

\bibitem{WallsMilb94}  D.~F.~Walls and G.~J.~Milburn,
        {\it Quantum Optics}, pages 92-97, (Springer, Berlin 1994).


\bibitem{Sun99} H.~B.~ Sun and G.~J.~Milburn, Phys Rev B,{\bf 59},
         10748, (1999).

%\bibitem{Gisin84} N. Gisin, Phys. Rev. Lett. {\bf 52}, 1657 (1984).

\bibitem{Dalibard92} J.~Dalibard, Y.~Castin, and K.~Molmer,
        Phys. Rev. Lett. {\bf 68}, 580 (1992).

\bibitem{Gisin92} N.~Gisin and I.~C.~Percival, J.~Phys. A {\bf 25} 5677 (1992);
         {\bf 26} 2233 (1993); {\bf 26} 2245 (1993).

\bibitem{Wiseman93}  H.~M.~Wiseman and G.J.~Milburn,
        Phys. Rev. A {\bf 47}, 1652 (1993).

\bibitem{Gagen93} M.~J.~Gagen, H.~M.~Wiseman, and G.~J.~Milburn,
        Phys. Rev. A {\bf 48}, 132 (1993).

\bibitem{Hegerfeldt93} G.~C.~Hegerfeldt, Phys. Rev. A {\bf 47}, 449 (1993).

\bibitem{Presilla96} C.~Presilla, R.~Onofrio, and U.~Tambini,
        Ann. Phys. {\bf 248}, 95 (1996).

\bibitem{Mensky98} M.~B.~Mensky, Phys. Usp. {\bf 41}, 923 (1998).


\bibitem{Plenio98} M.~B.~Plenio and P.~L.~Knight, Rev. Mod. Phys.
        {\bf 70}, 101 (1998).

\bibitem{Korotkov-derivation} After this manuscript was completed,
the authors received a preprint, Ref.\ \cite{Korotkov00c},
from Korotkov, in which a somewhat similar derivation to the authors'
approach for his Langevin rate equations is presented.

\bibitem{smallOmega} In obtaining Eq.\ (\ref{HI}), we have neglected in the
interaction picture the time dependence of the electron number
operator in dot 1 due to the tunneling term $\Omega$ in the CQDs,
$c_1^\dagger(t) c_1(t)\rightarrow c_1^\dagger c_1$.
This becomes exact when $\Omega=0$. This is reasonable if $\hbar
\sqrt{\varepsilon^2+4\Omega^2}\ll \max(|eV|,k_BT)$.
Here $\hbar \varepsilon=\hbar (\omega_2-\omega_1)$ is energy
mismatch between the two dots, $k_B$ is the
Boltzmann constant, $T$ represents the temperature,
$eV=\mu_L-\mu_R$ is the external bias applied across the PC,
and $\mu_L$ and
$\mu_R$ stand for the chemical potentials in the left and right
reservoirs respectively.
This assumption of small $\Omega$ or $\sqrt{\varepsilon^2+4\Omega^2}$,
made in Refs.\ \cite{Korotkov99,Korotkov00} and implicitly
in Ref.\ \cite{Gurvitz97}, can be understood as follows.
In deriving the master equation,
         we assume the electron tunneling amplitudes
         and density of states are almost
         constant over some bandwidth $\Delta\omega$ where tunneling may occur.
         Since the bandwidth $\hbar \Delta\omega$ is roughly in the
         order of magnitude of $\max(|eV|,k_BT)$,
         this is a good approximation if
         $|eV|, k_B T \ll \mu_{L(R)}$.
         We also assume the weak system-bath coupling, which implies
         the average electron tunneling rates,
         Eq.\ (\ref{rate}),
         $\hbar D, \hbar D' \ll \max(|eV|,k_BT)$. If $\Omega$, or more
         precisely the internal characteristic frequency
         of the CQD system $\sqrt{\varepsilon^2+4\Omega^2}$, is smaller than,
         or comparable to, $D,D'$, then it is all right to use Eq.\
         (\ref{HI}). On the other hand, if
         $\sqrt{\varepsilon^2+4\Omega^2}$ is much larger than $D,D'$,
         then we have to establish the condition to use Eq.\
         (\ref{HI}). In general, for finite $\Omega$, the electron
         tunneling through the PC may occur at different frequencies
         $\omega_k^L=\omega_k^R$ and $\omega_k^L=\omega_k^R \pm
         \sqrt{\varepsilon^2+4\Omega^2}$. In order for these frequency
         dependent electron tunneling amplitudes through the PC to be
         almost constant over the bandwidth $\Delta\omega$ as in the
         derivation for the master equation in the text,
         it is necessary to require
         $\hbar \sqrt{\varepsilon^2+4\Omega^2}\ll \max(|eV|,k_BT)$.
         If $\sqrt{\varepsilon^2+4\Omega^2}$ is not small enough as
         mentioned above, one instead has to
         include in Eq.\ (\ref{HI}) the time dependent phases
         $\exp(\pm i\sqrt{\varepsilon^2+4\Omega^2}\, t)$ and,
         beside the original term proportional to $c_1^\dagger c_1$,
         the dynamically generated terms via ${\cal H}_{CQD}$,
         such as terms proportional to
         $c_2^\dagger c_2$, $c_1^\dagger c_2$, and $c_2^\dagger c_1$.
         As a consequence, the electron tunneling rates through the PC
         Eq.\ (\ref{rate}), for example, may change
         and depend on the value of the internal characteristic frequency
         of the CQD system, $\sqrt{\varepsilon^2+4\Omega^2}$.


\bibitem{Ingold92} G.-L.~Ingold and Y.~V.~Nazarov,
        in {\it Single Charge Tunneling}, NATO ASI Series, Vol. B 294,
        edited by H.~Grabert and M.~H.~Devoret (Plem Press, New York, 1992).

\bibitem{Averin91} D.~A.~Averin and K.~K.~Likharev,
        in {\it Mesoscopic Phenomena in Solids}, edited by
        B.~L.~Alshuler, P.~A.~Lee, and
        R.~A.~Webb (Elsevier, Amsterdam, 1991).

\bibitem{Wiseman93b}  H.~M.~Wiseman and G.~J.~Milburn,
                       Phys. Rev. A {\bf 47}, 642 (1993).

\bibitem{Wiseman94}  H.~M.~Wiseman, Thesis,
                      (U. Queensland, Brisbane, Australia 1994)

\bibitem{Wiseman96} H.~M.~Wiseman, Quantum Semiclass. Opt. {\bf 8}, 
                    205 (1996).

\bibitem{Lindblad73} G.~Lindblad, Comm. Math. Phys. {\bf 48}, 119 (1973).

\bibitem{Larkin68} A.~I.~Larkin, Yu.~N.~Ovchinnikov, Sov. Phys. JETP
{\bf 26}, 1219 (1968); A.~J.~Dahm, A.~Denenstein, D.~N.~Langenberg,
W.H.~Parker, D.~Rogovin and D.~J.~Scalapino. Phys. Rev. Lett. {\bf 22}
1416 (1969); G.~Sch\"{o}n. Phys. Rev. B 32 4469 (1985).

\bibitem{Gardiner85} G.~W.~Gardiner, {\it Handbook of Stochastic
                     Method} (Springer, Berlin, 1985).

\bibitem{Oksendal92} B.~{\O}ksendal, {\it Stochastic Differential
        Equations} (Springer, Berlin, 1992).



\bibitem{mixingrate} The mixing rate for a qubit of a Josephson
        junction in the Coulomb-blockade regime, measured with a PC,
        is given in Ref.\ \cite{Makhlin00} as
        $\gamma_{\rm mix}=E_J^2 \tau_d /(1+{\Delta E}^2 \tau_d^2)$.
        There $E_J$ is the Josephson coupling, $\Delta E$ is the
        level spacing between the two logical states of the qubit,
        and $\tau_d$ is the decoherence time.
        In terms of the notations for the two-state CQD system
        considered here,
        $E_J \rightarrow 2\Omega$, $\Delta E \rightarrow \varepsilon$, and
        $\tau_d=(1/\Gamma_d)$, this mixing rate agrees with
        Eq.\ (\ref{mixrate}).
        In addition, the required condition
        $E_J \ll \max(\Delta E, \tau_d^{-1})$,
        under the above replacement rules, is consistent to
        that given in the text right below Eq.\ (\ref{mixrate}).





\end{references}
\end{document}